\documentclass[sigplan,nonacm]{acmart}

\newcommand{\ours}{TQC\xspace}
\usepackage{lipsum}
\usepackage{xspace}
\usepackage{booktabs}
\usepackage{tabularx}
\usepackage{makecell}
\usepackage{braket}
\usepackage{enumitem}
\usepackage{balance}

\usepackage{algorithm}
\usepackage{algpseudocode}

\begin{document}
\pagestyle{plain}

\title{Tableau-Based Framework for Efficient Logical Quantum Compilation}

\author{Meng Wang}
 \orcid{0009-0008-1749-7929}
 \affiliation{%
   \institution{Pacific Northwest National Lab \\ The University of British Columbia}
   \city{Richland}
   \country{USA}}
 \email{mengwang@ece.ubc.ca}

 \author{Chenxu Liu}
 \orcid{0000-0003-2616-3126}
 \affiliation{%
   \institution{Pacific Northwest National Lab}
   \city{Richland}
   \country{USA}}
 \email{chenxu.liu@pnnl.gov}

\author{Sean Garner}
 \orcid{0009-0008-1749-7929}
 \affiliation{%
   \institution{Pacific Northwest National Lab}
   \city{Richland}
   \country{USA}}
 \email{sean.garner@pnnl.gov}

 \author{Samuel Stein}
 \orcid{0000-0002-2655-8251}
 \affiliation{%
   \institution{Pacific Northwest National Lab}
   \city{Richland}
   \country{USA}}
 \email{samuel.stein@pnnl.gov}

 \author{Yufei Ding}
 \orcid{0000-0002-8716-5793}
 \affiliation{%
   \institution{University of California San Diego}
   \city{San Diego}
   \country{USA}}
 \email{yufeiding@ucsd.edu}

 \author{Prashant J. Nair}
 \orcid{0000-0002-1732-4314}
 \affiliation{%
   \institution{The University of British Columbia}
   \city{Vancouver}
   \country{Canada}}
 \email{prashantnair@ece.ubc.ca}
 
 \author{Ang Li}
 \orcid{0000-0003-3734-9137}
 \affiliation{%
   \institution{Pacific Northwest National Lab}
   \institution{University of Washington}
   \city{Richland}
   \country{USA}}
 \email{ang.li@pnnl.gov}

\renewcommand{\shortauthors}{Wang et al.}

\begin{abstract}

Quantum computing holds the promise of solving problems intractable for classical computers, but practical large-scale quantum computation requires error correction to protect against errors. Fault-tolerant quantum computing (FTQC) enables reliable execution of quantum algorithms, yet they often demand substantial physical qubit overhead. Resource-efficient FTQC architectures minimize the number of physical qubits required, saving more than half compared to other architectures, but impose constraints that introduce up to 4.7$\times$ higher runtime overhead. In this paper, we present \ours, a \underline{T}ableau-based \underline{Q}uantum \underline{C}ompiler framework that minimizes FTQC runtime overhead without requiring additional physical qubits. By leveraging operation reorderability and latency hiding through parallel execution, \ours reduces FTQC runtime overhead by \textbf{2.57$\times$} on average.

Furthermore, FTQC circuits often contain millions of gates, leading to substantial compilation overhead. To address this, we optimize the core data structure, the tableau, used in stabilizer formalism. We provide two tailored versions of the Tableau data type, each designed for different usage scenarios. These optimizations yield an overall performance improvement of more than \textbf{1000$\times$} compared to state-of-the-art FTQC optimization tools.

\end{abstract}

\maketitle %

\section{Introduction}
\label{sec:intro}

Quantum computing promises significant speedups on key problems including integer factorization~\cite{shor1999polynomial}, quantum chemistry simulation~\cite{lloyd1996universal}, and unstructured search~\cite{grover1996fast}. Yet today's quantum devices have high physical error rates, making fault tolerance essential. Quantum Error Correction (QEC) enables Fault-Tolerant Quantum Computation (FTQC), predominantly via stabilizer codes such as surface, color, and QLDPC~\cite{bravyi1998quantum,fowler2009high,fowler2012surface,Bombin2006, bombin2007exact,landahl2011fault,fowler2011two,babar2015fifteen,panteleev2021quantum,dinur2023good,panteleev2021degenerate,Bravyi2024}. To reduce the overall physical qubit footprint, several FTQC architectures expose only one logical basis (\(X\) or \(Z\)) per patch. This restriction forces \(Y\)-term decompositions and frequent basis realignments. This, in turn, adds serialization and significantly increases the runtime. To mitigate this, this paper aims to develop a compiler that reduces FTQC runtime \emph{without} adding physical qubits while keeping compilation scalable. %

\begin{figure}[t]
    \centering
    \includegraphics[width=0.9\linewidth]{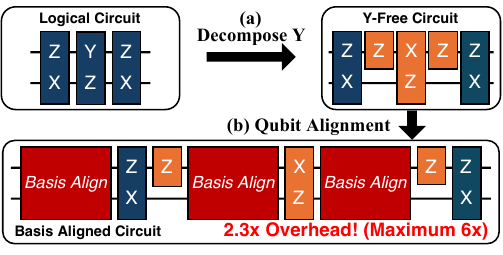}
    \caption{Logical quantum circuit in Pauli-product form and the additional steps required for FTQC execution. (a) Decomposition of all \(Y\)-basis operators into \(X\) and \(Z\) bases. (b) Basis alignment of logical qubits with the Pauli-product operator before application. These steps introduce an average 2.3$\times$ (upto a maximum of 6$\times$) FTQC runtime overhead.}
    \label{fig:ftqc_cost}
\end{figure}

An effective compiler needs an efficient program representation. In the stabilizer formalism~\cite{poulin2005stabilizer}, logical states and operations are tensor products of Pauli operators \((I, X, Y, Z)\). Thus, one can represent entire programs as ordered lists of Pauli products. Thus, as shown in Figure~\ref{fig:ftqc_cost}, one approach would be to use \emph{Pauli-product circuits (PPCs)} as an intermediate representation. PPCs make commutativity and basis requirements explicit, giving the compiler levers to reorder, fuse, and schedule operations while overlapping required basis rotations. Additionally,  PPCs are a concise, hardware-aware intermediate representation~\cite{litinski2019game, Azure_Quantum_Resource_Estimator} and serve as the program representation for analyses and transformations.

While PPCs make basis requirements explicit, FTQC is still constrained by the high physical qubit overhead. For example, the surface code uses nearly \(2d^2\) physical qubits per logical qubit~\cite{fowler2012surface}. Thus, several proposals use qubit-efficient, single-edge-access designs that expose only one logical basis (\(X\) or \(Z\)) per patch~\cite{litinski2019game}. This yields two direct requirements: (1) As shown in Figure~\ref{fig:ftqc_cost}a, any \(Y\) term must be rewritten as an \(X/Z\) product; and (2) as shown in Figure~\ref{fig:ftqc_cost}b, when a term targets the non-exposed basis, a basis rotation must be inserted before execution. These steps add operations and serialization, increasing the runtime by \(2.3\times\) (up to \(6\times\)) and typically undercut the qubit savings of single-edge access.

Prior solutions~\cite{litinski2019game, chamberland2022circuit} increase the physical footprint to expose multiple bases simultaneously. This usually requires $1.5\times$ to $4\times$ more qubits and limits the scalability of FTQC architectures. In contrast, this paper presents \ours, a \underline{T}ableau-based \underline{Q}uantum \underline{C}ompiler framework, that reduces runtime overhead on FTQC architectures while preserving their qubit savings. The \ours framework uses two key insights:

\noindent \textbf{1. Reordering through Commutativity:} Logical operations that commute can be reordered. For instance, the three logical operations in Figure~\ref{fig:ftqc_cost} commute, which allows their execution order to be swapped. By strategically reordering logical operations, \ours minimizes basis changes during execution, thereby reducing alignment overhead. It identifies maximal commuting subsets and schedules them to preserve program correctness while optimizing basis switching.

\noindent \textbf{2. Latency Hiding:} Basis alignment costs can be hidden by executing alignment operations concurrently with other logical gates. This overlap masks the latency of alignment by parallelizing it with useful computation on other qubits. Our scheduler detects such parallelism and prioritizes operations that can run alongside pending alignments.

\begin{figure}[t]
    \centering
    \includegraphics[width=\linewidth]{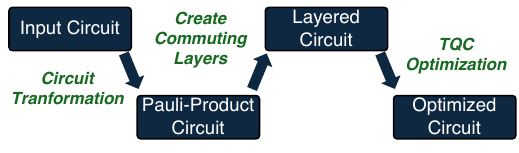}
    \caption{Flow of \ours optimization. The input circuit is first converted into a Pauli-product form. It is then organized into commuting layers, before applying \ours optimization.}
    \label{fig:opt_flow}
\end{figure}

Figure~\ref{fig:opt_flow} shows the overall optimization flow of \ours. The input circuit is first converted into a Pauli-product circuit, after which logical operations are grouped into commuting layers to expose reordering opportunities. \ours then applies its suite of optimizations, including gate reordering and latency hiding, ultimately achieving a 2.57$\times$ reduction in FTQC runtime overhead across a wide range of benchmarks.

While effective, \ours faces the same runtime challenges common to all FTQC optimization frameworks~\cite{heyfron2018efficient,Kissinger_2020,de2020fast,vandaele2024lowertcountfasteralgorithms,ruiz2025quantum}. These challenges stem mainly from the sheer number of gates in typical FTQC circuits, which often range from thousands to millions~\cite{gidney2021factor}. For example, a state-of-the-art T-gate optimizer takes over 14 hours to optimize a 16-qubit FTQC circuit with 71,700 gates~\cite{vandaele2024lowertcountfasteralgorithms}. Such excessive runtime makes these tools impractical for large-scale FTQC applications.

\noindent\textbf{Tableau for Efficient Compilation.} To make \ours scalable to large-scale FTQC circuits, we utilize a key data structure from stabilizer formalism: the tableau. A tableau is a compact binary representation of stabilizer groups that supports fast updates under Clifford operations using bitwise XOR operations only. It is widely used in quantum error correction frameworks, particularly stabilizer simulators~\cite{aaronson2004improved}, which simulate large circuits under Pauli noise and measurement to evaluate code performance and fault tolerance~\cite{Gidney_2021}.

\noindent\textbf{Limitations of Existing Tableau Designs.} While effective for simulation, tableaus are used very differently in circuit compilation. Simulators typically maintain a small tableau that is updated frequently, focusing on dynamic effects like noise injection and measurement collapse. In contrast, \ours requires constructing a much larger tableau—often 100$\times$ bigger—but does so only once, using it for static analysis of circuit structure. As a result, existing tableau implementations in simulators are not suitable for our setting.

To bridge this gap, \ours introduces custom tableau data structures tailored for circuit compilation. A key design challenge is how bits are stored within the tableau matrix. Since a tableau is binary, packing multiple bits together improves memory usage and computational efficiency. However, frequent access to individual bits can make unpacking costly. Tableau operations are used in two key stages shown in Figure~\ref{fig:opt_flow}: \textit{Circuit Transformation} and \textit{Creating Commuting Layers}. These stages have distinct data access patterns that guide how bits should be packed for optimal performance. To address this, we develop two specialized bit-packed tableau designs that eliminate costly bit operations during execution.

\noindent \textbf{1. Circuit Transformation:}
This stage primarily accesses tableau data column-wise, repeatedly querying bits corresponding to individual qubits. To optimize for this pattern, we introduce the \texttt{VTab} structure, which packs bits from multiple rows for each qubit together. This column-major packing enables simultaneous operations across many Pauli products for a single qubit, significantly improving efficiency.

\noindent \textbf{2. Creating Commuting Layers:}
In contrast, this stage requires frequent row-wise operations such as commutation checks and row reordering. We design the \texttt{HTab} structure to pack all qubit bits for each Pauli product row together. This row-major packing allows efficient bitwise operations on entire Pauli products without unpacking.

By carefully tailoring these two tableau layouts to their respective data access patterns, \ours achieves both high efficiency and strong scalability in the most computationally intensive compilation steps—without the overhead of bit unpacking. This optimized tableau design is a key factor enabling \ours to outperform existing DAG-based FTQC optimization frameworks by more than \textbf{1000$\times$} in speed.

This work makes the following contributions:
\begin{enumerate}[leftmargin=*]
    \item We tackle a major FTQC bottleneck, where hardware constraints increase runtime by 2.3$\times$ on average. 
    
    \item We present \ours, a compiler-level optimization that reduces FTQC runtime overhead by 2.57$\times$ across a wide range of benchmarks by leveraging commutativity-based reordering and latency hiding techniques.
    
    \item We introduce two specialized tableau data structures that enable efficient large-scale circuit compilation.
    
    \item We demonstrate that our tableau-based approach achieves over 1000$\times$ speedup compared to existing DAG-based FTQC optimization frameworks, making our compiler scalable to circuits with millions of gates.
    
    \item We evaluate \ours across diverse benchmarks, demonstrating significant runtime overhead reductions compared to state-of-the-art FTQC optimization frameworks, establishing its practicality for large-scale FTQC.
\end{enumerate}

\section{Background}
\label{sec:background}

\subsection{Quantum Error Correction and Stabilizer Codes}

QEC encodes logical qubits using multiple physical qubits to detect and correct errors without destroying the encoded quantum states~\cite{gottesman1997stabilizer,kitaev1997quantum}. QEC operates through syndrome extraction, where stabilizer measurements reveal error signatures without directly measuring the logical information. The stabilizer formalism~\cite{poulin2005stabilizer} provides the mathematical framework for this process, where a stabilizer code is defined by an abelian subgroup $S$ of the $n$-qubit Pauli group. The code space consists of states that are $+1$ eigenstates of all stabilizer generators, enabling systematic error detection through anticommutation relationships between errors and stabilizers.

\begin{figure}[h!]
    \centering
    \includegraphics[width=\linewidth]{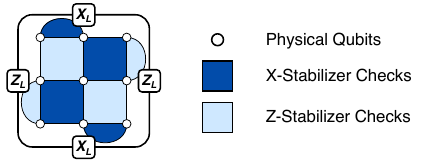}
    \caption{A distance-3 logical surface code patch with $X$-type (detecting $Z$ errors) and $Z$-type (detecting $X$ errors) stabilizers shown in different colors. Each stabilizer measures the parity of the qubits it touches. Logical $X$ and $Z$ operators span the patch between corresponding boundaries.}
    \label{fig:surface_code}
\end{figure}

The surface code is the most widely used QEC code for FTQC. Figure~\ref{fig:surface_code} shows a distance-3 code patch encoding one logical qubit. $X$- and $Z$-type stabilizers are shown in different colors, each measuring the parity of nearby qubits. Logical $X$ and $Z$ operators are defined as chains of Pauli operators along orthogonal edges of the patch. The logical $Y$ operator is the product of the logical $X$ and $Z$ operators. In QEC, one cycle is the execution of all stabilizer measurements once, while a QEC round consists of $d$ cycles and is used as the basic time unit for measuring FTQC execution time~\cite{litinski2019game}.

\subsection{Logical Operations}

FTQC relies on the universal Clifford+T gate set, comprising Pauli ($X$, $Y$, $Z$), Hadamard ($H$), Phase ($S$), controlled-$X$ (CNOT), and the non-Clifford $T$ gate~\cite{nielsen2010quantum}. The execution times for these gates are: $H$ and CNOT require $3d+4$ cycles, $S$ gates take $1.5d+3$ cycles, and $T$ gates require $2.5d+4$ cycles (where $T$ gate timing assumes magic states are pre-prepared)~\cite{blunt2024compilation}. Among these operations, CNOT, $S$, and $T$ gates require multi-qubit joint measurements: CNOT between two data qubits, while $S$ and $T$ gates use joint measurements for magic state injection with ancilla qubits~\cite{bravyi2005universal}.

\noindent \textbf{lattice Surgery:} The multi-qubit operations are implemented through lattice surgery~\cite{fowler2018low}, where the joint measurement requirements determine the logical edges involved. Lattice surgery works by merging and splitting code patches to realize joint operations. CNOT gates involve lattice surgery of $Z_1 \otimes X_2$ operators, requiring merging and splitting the $Z$ edge of the control qubit and the $X$ edge of the target qubit. Magic state injection for $S$ and $T$ gates requires $Z \otimes Z$ measurements, involving merging and splitting both $Z$ edges of the target qubit and ancilla. While transversal gates could alternatively be implemented on mobile qubit platforms like neutral atoms and trapped ions, qubit movement is approximately 1000$\times$ slower than gate operations on these platforms, making stationary lattice surgery the preferred approach.

\begin{figure}
    \centering
    \includegraphics[width=\linewidth]{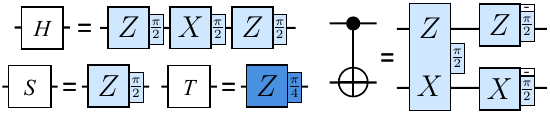}
    \caption{Logical operators expressed as Pauli-product operators. The main block indicates the qubit basis, while the side block shows the rotation angle and phase. }
    \label{fig:pp_ops}
\end{figure}

\noindent \textbf{Clifford+T as Pauli-Product Operations:} All gates in the Clifford+T set can be represented as Pauli-product operators~\cite{litinski2019game, Bravyi2016PBC,Peres2023PBC} as shown in Figure~\ref{fig:pp_ops}. This enables efficient compilation and optimization. Each Pauli-product operation is essentially a Pauli string representing which logical edge of which qubit is involved in the joint measurement. The $\pi/2$ rotations (such as $S$ gates) correspond to joint measurements between $|Y\rangle$ states and the target qubit's specified Pauli operator~\cite{gidney2024inplace}, taking $1.5d+3$ cycles each. The $\pi/4$ rotations ($T$ gates) involve joint measurements between $|T\rangle$ magic states and target qubits, requiring $2.5d+4$ cycles each. Gate commutation and transformations depend only on the Pauli strings and phases, independent of rotation angles.

\subsection{Resource-Efficient FTQC Architectures}

\begin{figure}[h!]
    \centering
    \includegraphics[width=\linewidth]{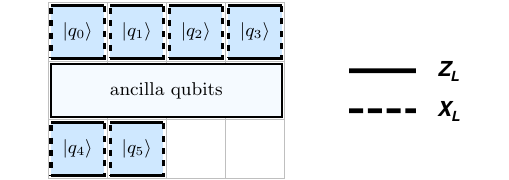}
    \caption{Resource-efficient FTQC architecture designed to save physical qubits. An ancillary region is reserved for non-local multi-qubit operations and patch rotations. Only one logical edge of each logical qubit patch is accessible.}
    \label{fig:compact_ftqc}
\end{figure}

Resource-efficient FTQC architectures are essential for near-term devices with limited physical qubits. A key constraint in such designs is that only one logical edge per qubit is accessible to reduce the spatial footprint and ancillary overhead, as shown in Figure~\ref{fig:compact_ftqc}. This necessitates logical patch rotations to access different edge types during operations, implemented via patch expansion and reconfiguration protocols that typically require about three QEC rounds. 

Pauli product operators acting on the $Y$ basis are particularly challenging because implementing $Y$ requires accessing both $X$ and $Z$ logical edges simultaneously. As a result, $Y$ operations must be decomposed into three sequential operations acting on the  $Z$, $X$, and $Z$ bases of a qubit. This increases the gate count and requires additional qubit patch rotations between consecutive operations.

\section{\ours: Logical Circuit Optimization}
\label{sec:design_ftqc_opt}

\begin{figure}[h!]
    \centering
    \includegraphics[width=\linewidth]{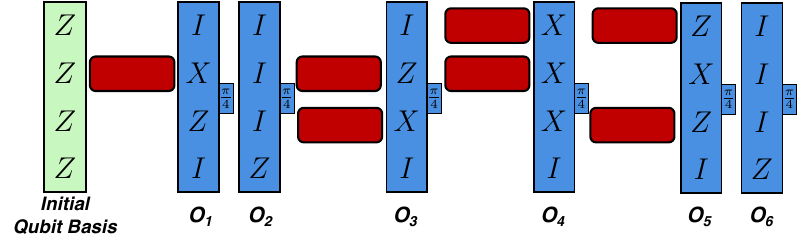}
    \caption{Baseline execution of 6 non-Clifford operations.}
    \label{fig:design_baseline}
\end{figure}

We begin by presenting the core optimization in \ours, which leverages the reorderability of logical Pauli-product operations and latency hiding to minimize FTQC runtime on resource-constrained architectures. Figure~\ref{fig:design_baseline} shows an example execution of 6 non-Clifford operations. Without any optimization, it takes 18 QEC rounds to complete.

\subsection{Commutation of Pauli-Product Operations}
\label{sec:pp_op_commutation}

The first step in \ours circuit optimization is identifying which Pauli-product operations can be reordered and determining an execution order that minimizes overall cost. Each operation is represented by a Pauli string specifying the basis operators on each qubit, with a sign bit for phase. For example, consider these Pauli strings on 4 logical qubits:

\[
+\, \underbrace{I}_{q_0} \underbrace{X}_{q_1} \underbrace{Z}_{q_2} \underbrace{I}_{q_3}
\]

This denotes a logical operation that involves the $X$ edge of qubit 1 and the $Z$ edge of qubit 3. Each Pauli-product operation consists of a Pauli string specifying the operators on each qubit and a rotation angle (e.g., $\frac{\pi}{2}$, $\frac{\pi}{4}$). The rotation angle is only relevant during physical implementation; for compilation and commutation analysis, we consider only the Pauli string. Commutation between two Pauli-product operations is determined by the pairwise commutation of their single-qubit components. The Identity operator ($I$) commutes with all Pauli operators, while distinct non-identity operators anticommute with each other:

\begin{equation}
    \begin{aligned}
&X Y = -Y X,\quad Y Z = -Z Y,\quad Z X = -X Z \\
\end{aligned}
\label{eq:anti_commutes}
\end{equation}
Given two Pauli strings P and Q:
\begin{equation}
P = P_1 \otimes P_2 \otimes \cdots \otimes P_n,\quad Q = Q_1 \otimes Q_2 \otimes \cdots \otimes Q_n,
\end{equation}
define the anticommutation indicator of P and Q at bit \( i \) as:
\begin{equation}
a_i = 
\begin{cases}
1, & \text{if } P_i \text{ and } Q_i \text{ anticommute} \\
0, & \text{otherwise}
\end{cases}
\label{eq:anticommute_indicator}
\end{equation}
Then, the commutation relationship between \( P \) and \( Q \) is:
\begin{equation}
PQ = QP \quad \Leftrightarrow \quad \sum_{i=1}^{n} a_i \equiv 0 \pmod{2}
\label{eq:commutation_condition}
\end{equation}

\subsection{Creating Commuting Layers}
\label{sec:layering}
Given a list of Pauli-product operators $\mathcal{O} = [O_1, \ldots, O_m]$, \ours first partitions them into commuting operation layers $\mathcal{L} = [L_1, \ldots, L_k]$, where operations within each layer can be freely reordered without affecting circuit semantics. A greedy approach would assign each operation to the most recent layer if it commutes with all operations there, otherwise creating a new layer. However, this can miss compatible earlier layers, resulting in unnecessary layer proliferation.

Instead, \ours uses an earliest-fit approach. Each operation is assigned to the earliest layer where it commutes with all existing operations, or a new layer is created if no such layer exists. This approach maximizes parallelization opportunities and minimizes the total number of layers. The layering algorithm is defined in Algorithm~\ref{alg:layering}.

\begin{algorithm}[h]
\caption{Create Commuting Layers with Earliest-Fit}
\label{alg:layering}
\begin{algorithmic}[1]
\Require List of FTQC operations \(\mathcal{O} = [O_1, \ldots, O_m]\)
\Ensure List of commuting layers \(\mathcal{L} = [L_1, \ldots, L_k]\)
\State Initialize \(\mathcal{L} \gets []\)
\For{each operation \(O_j\) in \(\mathcal{O}\)}
    \State \texttt{target\_layer} \(\gets -1\)
    \For{\(i = |\mathcal{L}| \text{ down to } 1\)}
        \If{\(O_j\) commutes with every operation in \(L_i\)}
            \State \texttt{target\_layer} \(\gets i\)
        \Else
            \State \textbf{break} \Comment{Stop checking earlier layers}
        \EndIf
    \EndFor
    \If{\texttt{target\_layer} \(\neq -1\)}
        \State Add \(O_j\) to \(L_{\texttt{target\_layer}}\)
    \Else
        \State Create new layer \(L_{\text{new}} \gets \{O_j\}\)
        \State Append \(L_{\text{new}}\) to \(\mathcal{L}\)
    \EndIf
\EndFor
\State \Return \(\mathcal{L}\)
\end{algorithmic}
\end{algorithm}
\begin{figure}[h!]
    \centering
    \includegraphics[width=\linewidth]{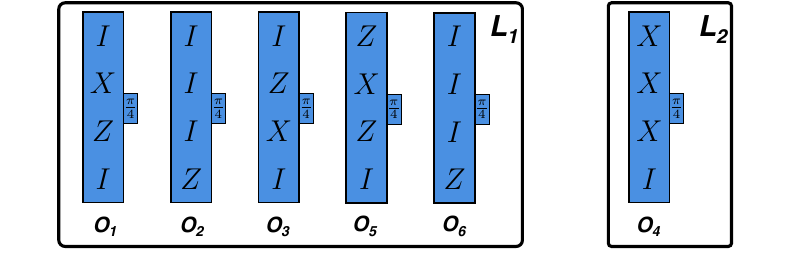}
    \caption{6 operations grouped into two commuting layers.}
    \label{fig:sample_layering}
\end{figure}
Compared to the \textit{greedy} method, the \textit{earliest-fit} algorithm significantly improves layering efficiency. For example, in a 30-qubit Quantum Fourier Transform circuit, \textit{earliest-fit} increases average operations per layer by over seven times and reduces the total number of layers from 80,386 to 11,252, greatly enhancing optimization potential downstream. Figure~\ref{fig:sample_layering} shows this layering process with 6 operations partitioned into two commuting layers.

\subsection{Operation Fusion}
\label{subsec:grouping}
After partitioning operations into commuting layers, \ours performs a second optimization: fusing operations within each layer that share identical Pauli strings. Since operations in the same layer commute, those acting on the identical qubits with the same Pauli operators can be combined into a single, more efficient fused operation.

\subsubsection{Fusion Process}
\begin{figure}[h!]
    \centering
    \includegraphics[width=\linewidth]{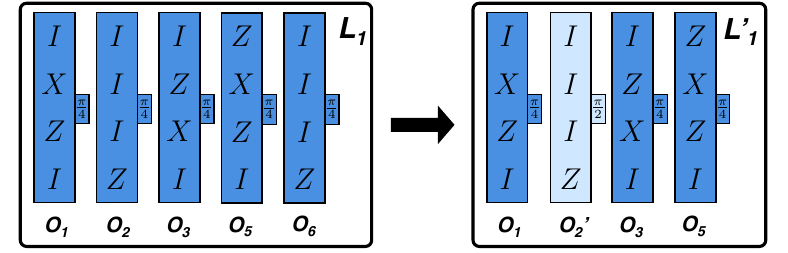}
    \caption{Operation fusion example for layer $L_1$. Non-Clifford operations $O_2$ and $O_6$ with identical Pauli strings are fused into a single Clifford operation $O_2'$.}
    \label{fig:fusion_example}
\end{figure}

Operations with identical Pauli strings are merged by summing their rotation angles:
\begin{equation}
    \theta_{\text{fused}} = \sum_{k=1}^{n} \theta_k
    \label{eq:total_angle}
\end{equation}
where $\theta_k$ represents each operation's signed rotation angle. Figure~\ref{fig:fusion_example} demonstrates this process using operations from layer $L_1$ in the previous example, where non-Clifford $O_3$ and $O_4$ are fused into a single Clifford $O_3'$. 

\subsubsection{Cost Reductions}
Operation fusion yields significant cost savings through three mechanisms:

\noindent \textbf{Complete elimination:} When fused angles sum to zero ($\theta_{\text{fused}} = 0$), operations cancel entirely.

\noindent \textbf{Gate simplification:} Fused angles that equal standard rotations become cheaper to execute. Non-Clifford operations ($\frac{\pi}{4}$) can combine into Clifford gates ($\frac{\pi}{2}$), or Clifford gates can combine into Pauli operations ($\pi$), which have zero cost.

\noindent \textbf{Execution efficiency:} When fusion produces non-standard angles, the merged operation can be executed at the cost of the most expensive original operation. For example, fusing $\frac{\pi}{4}$ and $\frac{\pi}{2}$ operations yields a $\frac{3\pi}{4}$ rotation that executes with the cost of a single $\frac{\pi}{4}$ gate, saving the $\frac{\pi}{2}$ operation cost.

Through this fusion process, \ours substantially reduces circuit complexity and execution cost.

\subsection{Minimizing Basis Alignment}
\label{subsec:reordering}
After operation fusion, \ours optimizes the execution order within each layer to minimize costly basis changes between operations. In FTQC, each logical qubit has an accessible logical operator (either Z or X) called its "exposed basis." When a Pauli operation requires an inaccessible logical operator, the code patch must be rotated before the gate can be applied. This code patch rotation requires 3 QEC rounds, which is a significant overhead since regular logical operations complete in only 1 QEC round.

\subsubsection{Basis-Aware Ordering}
\ours uses a simple heuristic to minimize basis rotations:
\begin{enumerate}[leftmargin=*]
    \item Start with all qubits exposing their Z basis
    \item From remaining operations in the current layer, select the one requiring the fewest code patch rotations based on current exposed bases
    \item Execute the selected operation and update each affected qubit's exposed basis accordingly  
    \item Repeat until all operations in the layer are complete
\end{enumerate}

\begin{figure}[h!]
    \centering
    \includegraphics[width=\linewidth]{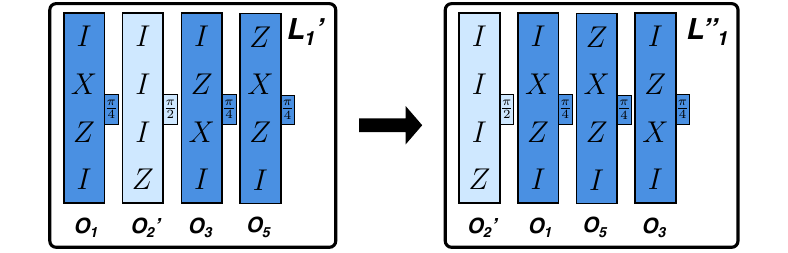}
    \caption{Basis-aware reordering example. The original operation sequence $O_1, O_2', O_3, O_5$ in $L_1'$ is reordered to $O_2', O_1, O_5, O_3$ to minimize code patch rotations by selecting operations that match the current exposed basis.}
    \label{fig:basis_example}
\end{figure}

The final basis configuration from each layer becomes the starting state for the next layer, ensuring basis optimizations compound across the entire circuit. Figure~\ref{fig:basis_example} demonstrates this reordering process where in layer $L_1$ the original sequence $O_1, O_3', O_4$ is reordered to $O_3', O_4, O_1$.

This reordering aims to reduce basis transformation overhead, which constitutes a major cost component in resource-constrained FTQC architectures as shown in Figure~\ref{fig:ftqc_cost}.

\subsection{Hiding Basis Rotation Latency}
\label{subsec:concurrent_basis_rotation}

Even after basis-aware reordering, some code patch rotations remain unavoidable. \ours addresses this by proactively scheduling these rotations to overlap with other operations, effectively hiding their 3 QEC round latency.

The optimization works by analyzing the reordered operation sequence to identify all necessary basis changes, then scheduling each rotation at the earliest possible moment before its dependent operation. While a qubit undergoes basis rotation, operations on other qubits can execute concurrently, utilizing the available parallelism within each layer.

\begin{figure}[h!]
    \centering
    \includegraphics[width=\linewidth]{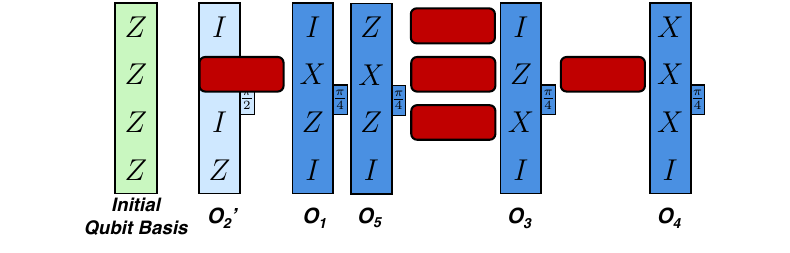}
    \caption{Optimized execution through scheduling techniques. FTQC execution time reduced to 13 QEC rounds.}
    \label{fig:latency_hiding}
\end{figure}

Figure~\ref{fig:latency_hiding} shows the optimized scheduling of the same operations shown in Figure~\ref{fig:design_baseline}. Through a combination of operation layering, merging, reordering, and parallel basis rotations, \ours reduces the overall FTQC execution time from 18 QEC rounds to 13 QEC rounds in this example.

\section{\ours: Tableau-Based Compilation}
\label{section:design_runtime}

Traditional quantum circuit compilation relies on DAG (Directed Acyclic Graph) approaches, which often introduce significant runtime overhead due to irregular memory access patterns and poor cache performance. The node connections in a DAG depend on the input circuit's gate configuration, resulting in inefficient processing for large circuits.

To demonstrate this limitation, we compared our tableau-based approach with Qiskit (which uses the high-performance Rust-based graph library rustworkx) on a 30-qubit QFT circuit containing 207,594 gates:

\begin{table}[h]
\centering
\caption{Performance comparison on a 30-qubit QFT circuit}
\begin{tabular}{lcc}
\hline
\textbf{Approach} & \textbf{Task} & \textbf{Runtime} \\
\hline
Qiskit (DAG-based) & Basic gate merging & 290s \\
\ours (Tableau-based) & Full optimization suite & 0.8s \\
\hline
\end{tabular}
\label{tab:compilation_performance}
\end{table}

As shown in Table~\ref{tab:compilation_performance}, \ours achieves a 363$\times$ speedup while performing more advanced optimizations, including gate commutation, layering, and reordering. This significant improvement is enabled by the tableau data structure’s regular memory layout and vectorized operations, which enable efficient circuit representation and cache-friendly computation.

\subsection{Tableau and Stabilizer Simulator}

\begin{figure}[h!]
    \centering
    \includegraphics[width=0.95\linewidth]{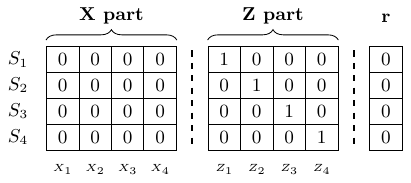}
    \caption{Tableau for a 4-qubit system in state $|0000\rangle$. Each row contains X bits, Z bits, and a phase bit $r$. }
    \label{fig:tableau}
\end{figure}

The stabilizer tableau encodes $n$ stabilizer generators for an $n$-qubit system as a binary matrix, as shown in Figure~\ref{fig:tableau}. Each row $S_i$ contains an X part ($n$ bits for Pauli X operators), a Z part ($n$ bits for Pauli Z operators), and a phase bit $r$ that determines the sign $(-1)^r$ of the Pauli operator.

The bit pair $(X_j, Z_j)$ encodes the Pauli operator on qubit $q_j$: $(0,0) \rightarrow I$, $(1,0) \rightarrow X$, $(0,1) \rightarrow Z$, and $(1,1) \rightarrow Y$. Thus each stabilizer represents a Pauli string $S_i = (-1)^{r_i} P_1 \otimes P_2 \otimes \cdots \otimes P_{n}$ where $P_j \in \{I, X, Y, Z\}$ is determined by the corresponding bit pair in row $i$.

The stabilizer tableau can efficiently simulate an n-qubit Clifford circuit by maintaining $n$ stabilizer generators in $n$ rows, requiring only $n \times (2n+1)$ bits total. This yields $O(n^2)$ complexity, compared to exponential $O(2^n)$ scaling in general state vector simulations. 

Simulation begins by initializing each qubit with its $Z$ stabilizer, placing the system in the computational $Z$ basis state, as shown in Figure~\ref{fig:tableau}. Each row of the tableau corresponds to a stabilizer generator and contains $X$ bits, $Z$ bits, and a phase bit $r$. When a Clifford gate is applied, the simulator iterates through all the rows and updates the relevant bits:

\begin{itemize}[leftmargin=*]
  \item \textbf{Single-qubit gate:} the $x_i$, $z_i$ bits of the target qubit (index $i$), and the phase bit $r$, are updated in each row.
  
  \item \textbf{CNOT gate:} the $x_i$, $z_i$ bits of the control qubit (index $i$), the $x_j$, $z_j$ bits of the target qubit (index $j$), and the phase bit $r$, are updated in each row.
\end{itemize}

This update process is entirely row-local: each stabilizer is modified independently, without reference to other rows.

\subsection{Tableau for Pauli-Product Circuit Compilation}

While the runtime bottleneck of stabilizer simulation is primarily measurements~\cite{Gidney_2021}, which require cross-row operations and probabilistic noise tracking, Pauli-product circuit compilation presents entirely different challenges. Instead of operating on a fixed $n$-row tableau, compilation works with much larger tableaux containing thousands or millions of rows. However, unlike simulation, these tableaux do not require repeated modifications during processing.

\ours uses tableaux in two main components: preprocessing Pauli-product circuits and checking commutation relationships between Pauli product operators.

\noindent \textbf{Preprocessing:} In FTQC, a common optimization commutes all Clifford gates to the circuit end and merges them into measurements~\cite{litinski2019game}. This process, traditionally relying on DAGs, can be performed much more efficiently using Tableau.

\noindent \textbf{Commutation check:} For two Pauli product operators represented as stabilizer rows $S_1$ and $S_2$, the commutation relationship from Section~\ref{sec:pp_op_commutation} reduces to:

\begin{equation}
 ((S_1.x \land S_2.z) \oplus (S_1.z \land S_2.x)) \land 1 == 0   
 \label{eq:tab_commute_check}
\end{equation}
This bitwise operation is significantly more efficient than iterating through bit pairs from two Pauli strings, counting anticommuting terms, and checking parity.

These two applications of Tableau involve distinct data access patterns. To accommodate both efficiently, \ours introduces two tailored tableau structures: \textbf{VTab} and \textbf{HTab}.

\subsection{VTab for Circuit Preprocessing}
\label{sec:pbc}

The Clifford commuting process tracks two key aspects: how non-Clifford Pauli operators transform as Clifford gates commute through them, and how final measurements appear after merging Clifford gates into them. 

A T gate can be represented as a Pauli-Z operator on the target qubit. When commuting a Clifford gate through it, the resulting T operator is given by applying the Clifford to the original Z operator. Measurement can also be modeled using a Z operator, since we measure in the computational basis.

\begin{figure}[h!]
    \centering
    \includegraphics[width=\linewidth]{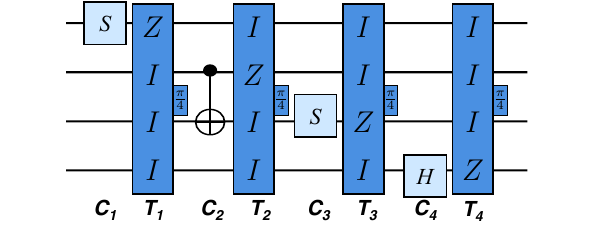}
    \caption{Example circuit with 4 Clifford and 4 T gates.}
    \label{fig:pbc_pre}
\end{figure}

The challenge is tracking which Clifford gates apply to which T gate stabilizers. Consider the example circuit in Figure~\ref{fig:pbc_pre} with 4 Clifford gates and 4 T gates. Gate T1 should have only C1 applied to it, T2 should have C2 and C1 applied in that order, T3 should have C3, C2, and C1 applied in that order, and T4 plus all final measurements should have all Clifford gates applied in the order C4, C3, C2, C1. A convenient approach initializes a tableau similar to Figure~\ref{fig:tableau} and traverses the circuit from back to front. When encountering a T gate, we add one stabilizer row to the tableau. When encountering a Clifford gate, apply it to the entire tableau.

\begin{figure}[h!]
    \centering
    \includegraphics[width=0.95\linewidth]{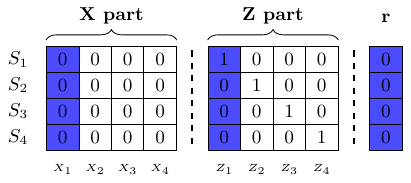}
    \caption{Tableau cells involved when applying a single-qubit Clifford gate on the first qubit. Highlighted columns show the vertical data access pattern that motivates VTab's column-wise bit-packing design.}
    \label{fig:vtab_demo}
\end{figure}

When applying Clifford gates to the tableau, the operation iterates through all stabilizers and accesses the X and Z bits of the target qubit plus the phase bit for each stabilizer. Figure~\ref{fig:vtab_demo} illustrates this by highlighting the tableau cells involved when applying a single-qubit Clifford gate to the first qubit. The highlighted pattern reveals that entire columns are accessed together, creating a vertical data access pattern.

This access pattern is ideal for packing columns into 64-bit unsigned integers (the largest integer type on 64-bit systems), where each integer contains 64 rows of a single column. Bitwise operations can then process 64 rows in parallel with a single operation, significantly accelerating gate applications. Since entire columns are accessed for each gate operation, we store consecutive packed data for each column together to optimize cache performance. Since this design packs column data vertically into dense data types, we call it VTab.

\subsubsection{Trade-offs in Clifford Gate Preprocessing}
While the standard preprocessing approach pushes and merges all Clifford gates into measurements~\cite{litinski2019game}, this introduces several trade-offs. First, as shown in Figure~\ref{fig:pp_ops}, the default Clifford+T gate set operates entirely in $X$ and $Z$ bases, but commuting Clifford gates through $T$ gates converts some qubits to the $Y$ basis. As shown in Figure~\ref{fig:ftqc_cost}, such Y-basis operations are very costly to run. Second, when pushing two-qubit $Z \otimes X$ $\pi/2$ rotations through $T$ gates, the Pauli weight of $T$ gates increases if they share exactly one qubit with the CNOT gate. This weight increase transforms the original single-qubit $Z$ Pauli strings of $T$ gates into multi-qubit $\pi/4$ rotations, potentially reducing circuit parallelism. \ours monitors these changes and selectively keeps Clifford gates in place when commutation would hurt performance.

\subsubsection{Parallelization with MPI} 
\label{sec:parallel_pbc}Clifford operations are embarrassingly parallel—each stabilizer row can be transformed independently without dependencies between stabilizer rows. We implement a parallelized VTab using MPI. We choose MPI rather than shared-memory approaches like OpenMP because we want each process to have its own dedicated memory space. This ensures optimal cache performance for our carefully designed, packed data structures.

Stabilizer rows are partitioned equally across processes using round-robin distribution to maintain load balancing as stabilizer rows are added sequentially:

\begin{center}
\begin{tabular}{c|c|c|c}
\textbf{Process 1} & \textbf{Process 2} & \textbf{Process 3} & \textbf{Process 4} \\
\hline
$S_1$ & $S_2$ & $S_3$ & $S_4$ \\
$S_5$ & $S_6$& $S_7$ & $S_8$ \\
$S_9$ & $S_{10}$ & $S_{11}$ & $S_{12}$ \\
$S_{13}$ & $S_{14}$ & $S_{15}$ & $S_{16}$ \\
$\vdots$ & $\vdots$ & $\vdots$ & $\vdots$ \\
\end{tabular}
\end{center}

This distribution pattern ensures a balanced workload regardless of the number of stabilizers currently active.  

\subsubsection{Precomputed Stabilizer Rows for Toffoli}

The 3-qubit CCX (Toffoli) gate is essential in FTQC, appearing frequently in important algorithms including Shor's algorithm~\cite{shor1999polynomial} and quantum phase estimation~\cite{Babbush_2018}. A single Toffoli gate decomposes into 7 T gates plus 8 Clifford gates~\cite{nielsen2010quantum}.

The standard approach would sequentially add T gates and apply Clifford gates as previously described. However, we observe that the 8 Clifford gates, when combined, result in the identity operation. This means that after processing a Toffoli gate, the original stabilizer rows in the tableau remain unchanged—the only effect is adding 7 new stabilizer rows.

Since these 7 rows depend only on the target qubit indices, they can be precomputed and inserted directly when encountering a Toffoli gate. \ours implements this optimization to reduce the execution time significantly.

\subsection{HTab for Efficient Layering and Reordering}

After eliminating Clifford gates during preprocessing, the next step performs \ours layering and optimizations from Section~\ref{sec:design_ftqc_opt}. The most computationally expensive part is creating commuting layers, which requires commutation checks between Pauli operators. These operations have a completely different data access pattern from Clifford gate applications.

\begin{figure}[h!]
    \centering
    \includegraphics[width=0.93\linewidth]{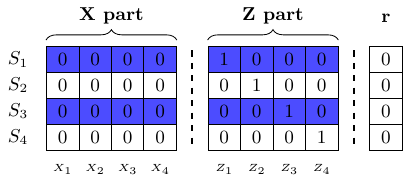}
    \caption{Data access pattern when checking commutation between stabilizers $S_1$ and $S_2$. Highlighted cells show the row-wise access pattern that motivates HTab's design.}
    \label{fig:htab_demo}
\end{figure}

Figure~\ref{fig:htab_demo} illustrates the data access pattern when checking commutation between stabilizers $S_1$ and $S_2$. Unlike Clifford operations that access entire columns, commutation checks access complete rows of data. Using VTab for these operations would incur significant overhead from bit packing and unpacking with masking, since VTab's column-wise organization is optimized for column-wide gate applications.

To address this, we design HTab, which packs the X and Z parts of each row horizontally together into dense data types. Since commutation checks require both X and Z components of two rows (as shown in Equation~\ref{eq:tab_commute_check}), storing them contiguously improves cache locality. 

During the layering operation, each layer stores its stabilizers in an HTab sorted by descending Pauli weight (number of non-identity Pauli operators). When determining if a candidate operator can join a layer, we check compatibility with existing operators in weight order. Since higher-weight stabilizers are more likely to anticommute with the candidate, this ordering enables early termination. We can reject the candidate as soon as we find any anticommuting pair, avoiding unnecessary checks with the remaining operators.

Weight-based sorting occurs only during initial layer creation. Once established, subsequent optimizations like gate fusion and basis alignment reordering do not preserve this ordering. Nevertheless, the initial weight-based organization benefits from gate fusion since identical Pauli strings necessarily have identical weights. This allows fusion operations to efficiently target contiguous weight groups within each layer rather than scanning the entire layer.

\section{Evaluation Methodology}
\label{sec:methodology}

\subsection{Figure of Merit}

The primary figure of merit is execution time measured in QEC rounds required to complete a given FTQC circuit on the resource-efficient architecture shown in Figure~\ref{fig:compact_ftqc}. Each QEC round corresponds to d error correction cycles, where d is the code distance of the QEC code.

We developed a round-accurate simulator that models the execution constraints of this architecture. This simulator schedules quantum operations according to circuit dependencies while tracking the accessibility of logical qubit edges required for interactions. When necessary, edges are not exposed; the simulator inserts qubit rotation operations, each taking 3 rounds to complete. The simulator carefully manages ancilla qubit allocation, ensuring each ancilla is dedicated to a single operation at a time and appropriately stalls operations when insufficient ancilla qubits are available.

\subsection{Benchmark Circuits}

We evaluate \ours using a diverse set of FTQC benchmarks, including Quantum Fourier Transform (QFT), Quantum Phase Estimation (QPE), Ising Model simulations, and W-state preparation~\cite{li2022qasmbenchlowlevelqasmbenchmark}. We also incorporate key arithmetic benchmarks from the Op-T-mize dataset~\cite{Kottmann2024_op-T-mize}, which are primarily Toffoli-heavy circuits representing essential subroutines in major FTQC algorithms such as Shor's algorithm~\cite{shor1999polynomial}, Grover's search~\cite{grover1996fast}, quantum chemistry simulations~\cite{Babbush_2018}, and the HHL algorithm~\cite{harrow2009quantum,childs2017quantum} for linear systems.

\noindent\textbf{Benchmark Preprocessing} 
All benchmark circuits are first converted to Clifford+T form. For benchmarks with arbitrary rotations (e.g., $R_z$), we use the GridSynth algorithm~\cite{ross2014optimal} with a default synthesis error of $10^{-10}$. The resulting Clifford+T circuits are then mapped to Pauli product operations, after which Clifford gates are removed using the method described in Section~\ref{sec:pbc}. The resulting Clifford-free circuits serve as the baseline for fair comparison. Details of each benchmark circuit are summarized in Table~\ref{tab:benchmark_results}.

\subsection{Optimization Runtime Comparison}

We compare \ours with the fastest state-of-the-art FTQC T-gate optimization frameworks, PHAGE~\cite{de2020fast} and TOHPE~\footnote{Other methods exist, but these two remain the most efficient in practice.}. All methods use the Op-T-mize benchmark suite, but prior works often omitted runtime results (either exceeding 24 hours or under 1 second). We therefore focus on four benchmarks with reported non-trivial runtimes: \texttt{ham15-high}, \texttt{hwb8}, \texttt{hwb10}, and \texttt{mod\_adder\_1024}. For these, we report our speedup relative to the faster of PHAGE and TOHPE. Even on the largest benchmarks, \ours finishes in under a minute, whereas prior works require over 24 hours.

For Clifford gate elimination preprocessing (Section~\ref{sec:pbc}), we compare against PennyLane's optimization pass. Since PennyLane proved too inefficient for our smallest benchmarks, we generated smaller random circuits for comparison: 20-qubit circuits with 70-150 gates, of which half are Clifford gates and half are non-Clifford gates.

\subsection{Hardware Configuration}

To provide a fair and grounded runtime comparison, we conducted all evaluations on a laptop with a 10-core ARM processor running at 3.2 GHz. By using standard consumer hardware instead of a high-performance computing cluster, we ensure that any observed performance advantages stem directly from the efficiency of our algorithm, not from access to superior hardware. This also highlights the practicality of \ours, demonstrating its effectiveness and accessibility for users without specialized computing infrastructure. For MPI implementation, we use OpenMPI 5.0.7.

\section{Results}

\subsection{FTQC Execution Speedup}

Table~\ref{tab:benchmark_results} shows the execution time improvement of \ours across a diverse set of benchmarks. For each circuit, we show the number of qubits, total gates, baseline FTQC execution cycles, \ours optimized execution cycles, and the resulting speedup. Overall, \ours consistently improves execution time, achieving an \textbf{average speedup of 2.57$\times$} across all benchmarks. The speedup ranges from \textbf{1.37$\times$} on small arithmetic circuits to over \textbf{12$\times$} on structured benchmarks with deep circuit depth such as \texttt{qft\_n18}. 

\begin{table}[htbp]
\centering
\caption{\ours circuit execution speedup}
\label{tab:benchmark_results}
\small
\begin{tabular}{@{}l@{\hspace{7pt}}c@{\hspace{7pt}}c@{\hspace{7pt}}c@{\hspace{7pt}}c@{\hspace{7pt}}c@{}}
\toprule
\textbf{Circuit} & \textbf{Qubits} & \textbf{Gates} & \textbf{Baseline} & \textbf{TQC} & \textbf{Speedup} \\
 & & & \textbf{Cycles} & \textbf{Cycles} & \\
\midrule
adder\_8 & 24 & 900 & 633 & 406 & 1.56 \\
barenco\_tof\_10 & 19 & 450 & 630 & 358 & 1.76 \\
csla\_mux\_3 & 15 & 170 & 138 & 101 & 1.37 \\
cycle\_17\_3 & 35 & 9488 & 12357 & 8000 & 1.54 \\
grover\_5 & 9 & 789 & 1107 & 627 & 1.77 \\
ham15-med & 17 & 1206 & 1900 & 1067 & 1.78 \\
ham15-high & 20 & 4966 & 6086 & 3787 & 1.61 \\
hwb8 & 12 & 14382 & 42329 & 22716 & 1.86 \\
hwb10 & 16 & 71700 & 193438 & 75163 & 2.57 \\
ising\_n26 & 26 & 18995 & 17026 & 3314 & 5.14 \\
mod\_adder\_1024 & 28 & 4045 & 4980 & 2942 & 1.69 \\
mod\_red\_21 & 11 & 268 & 327 & 204 & 1.60 \\
qcla\_adder\_10 & 36 & 521 & 162 & 118 & 1.37 \\
qcla\_com\_7 & 24 & 441 & 180 & 123 & 1.46 \\
qcla\_mod\_7 & 26 & 884 & 585 & 309 & 1.89 \\
qft\_n18 & 18 & 96693 & 623362 & 50985 & \textbf{12.23} \\
qpe\_n9 & 9 & 7577 & 33839 & 14485 & 2.34 \\
rc\_adder\_6 & 14 & 200 & 801 & 182 & 4.40 \\
tof\_10 & 19 & 255 & 360 & 177 & 2.03 \\
wstate\_n76 & 76 & 38214 & 89404 & 59478 & 1.50 \\
\bottomrule
\end{tabular}
\end{table}

These results highlight two key observations: (1) Even modestly sized benchmarks exhibit non-trivial speedups, demonstrating that \ours consistently reduces FTQC execution cycles. (2) Circuits with large depth or heavy Clifford structure (e.g., \texttt{qft\_n18}) see order-of-magnitude improvements, showing that \ours scales exceptionally well with increasing circuit complexity. In summary, \ours achieves substantial reductions in FTQC execution time across a broad range of benchmarks, with improvements of up to \textbf{12.23$\times$}, confirming its effectiveness and scalability.

\subsection{\ours Runtime Overhead}

\begin{table}[htbp]
\centering
\caption{\ours Optimization Time}
\label{tab:tqc_times}
\small
\begin{tabular}{@{}lc|@{\hspace{15pt}}lc@{}}
\toprule
\textbf{Circuit} & \textbf{Time (ms)} & \textbf{Circuit} & \textbf{Time (ms)} \\
\midrule
adder\_8        & 4.4       & mod\_adder\_1024 & 10.6     \\
barenco\_tof\_10 & 1.5      & mod\_mult\_55    & 0.17      \\
csla\_mux\_3    & 0.6       & mod\_red\_21     & 0.3      \\
cycle\_17\_3    & 70.5      & qcla\_adder\_10  & 1.4      \\
grover\_5       & 1.0       & qcla\_com\_7     & 0.6      \\
ham15-med       & 1.9       & qcla\_mod\_7     & 1.5      \\
ham15-high      & 13.4      & qft\_n18         & 26653  \\
hwb8            & 3485    & qpe\_n9          & 43     \\
hwb10           & 3488    & rc\_adder\_6     & 0.46      \\
ising\_n26      & 481     & tof\_10          & 0.9      \\
mod5\_4         & 0.09       & wstate\_n76      & 9487   \\
\bottomrule
\end{tabular}
\end{table}

FTQC algorithms are typically large in scale, often comprising thousands of gates, as shown in Table~\ref{tab:benchmark_results}. Without an efficient optimization framework, FTQC compilation itself becomes impractical, regardless of the quality of the resulting circuits. Table~\ref{tab:tqc_times} reports the time taken by \ours to perform full optimization for each benchmark circuit. The runtime generally scales with the number of gates, with most circuits completing in just a few milliseconds — a notable achievement compared to existing general-purpose FTQC optimization frameworks, which we will compare against in Section~\ref{sec:runtime_vs_t_opt}. Even the largest benchmark in our suite, the 18-qubit QFT, completes in under 27 seconds.

\subsection{Runtime Overhead vs.~General Optimization}
\label{sec:runtime_vs_t_opt}

Since no existing framework performs the exact optimization procedure as \ours, a direct comparison of runtime overhead is not possible. Instead, we compare \ours against state-of-the-art general FTQC optimization frameworks that are known for their runtime performance. These frameworks primarily focus on reducing the number of $T$ gates, employing techniques that traverse the circuit to analyze gate relationships, merge operations, and eliminate redundancies — a process broadly similar to that of \ours. However, a key difference lies in the underlying representation: while existing tools typically operate on a DAG representation of the circuit, \ours performs its optimization using a Tableau-based approach.

\begin{figure}[h!]
    \centering
    \includegraphics[width=\linewidth]{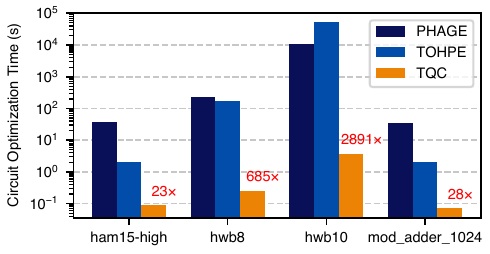}
    \caption{Circuit optimization runtime comparison between PHAGE, TOHPE, and \ours. \ours achieves up to \textbf{2891$\times$ speedup}, reducing optimization time from hours to seconds.}
    \label{fig:ftqc_speedup}
\end{figure}

Figure~\ref{fig:ftqc_speedup} compares the circuit optimization runtime of \ours against state-of-the-art $T$-gate optimization frameworks PHAGE and TOHPE on representative \texttt{Op-T-mize} benchmarks. We observe that \ours consistently achieves \textbf{orders of magnitude faster optimization}. For example, on \texttt{hwb10}, PHAGE and TOHPE require over $10^4$ to $10^5$ seconds, while \ours completes in under 3.5 seconds, yielding a 2891$\times$ speedup. Even for smaller circuits such as \texttt{ham15} and \texttt{mod\_adder}, \ours is more than an order of magnitude faster, achieving 23$\times$ and 28$\times$ speedups, respectively. On \texttt{hwb8}, we observe a 685$\times$ speedup. Importantly, all benchmarks in Table~\ref{tab:benchmark_results} that previously required hours or even days to optimize can now be handled by \ours in under 30 seconds. This demonstrates that our approach removes the runtime bottleneck present in prior $T$-gate optimization frameworks, enabling practical scalability to larger quantum circuits.

\subsection{Runtime Overhead vs.~PennyLane}

Another vital aspect of runtime comparison involves the preprocessing step that converts circuits to non-Clifford-only form, as discussed in Section~\ref{sec:pbc}. This preprocessing is a general optimization employed by many FTQC frameworks to simplify subsequent steps. Such functionality is also provided by PennyLane, making it a natural baseline for comparison. Since PennyLane could not handle our larger benchmark circuits in a reasonable time, we created smaller random circuits with 20 qubits and between 100 and 290 gates. About half of the gates in each circuit are Clifford gates, and the other half are non-Clifford (specifically $T$ gates).

\begin{figure}[h!]
    \centering
    \includegraphics[width=\linewidth]{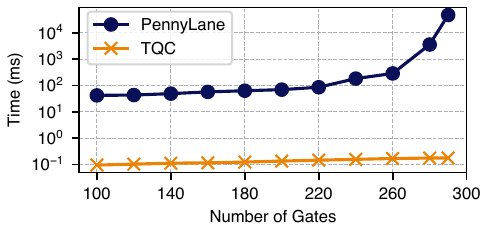}
    \caption{Clifford gate elimination runtime comparison between PennyLane and \ours on random 20-qubit circuits. \ours runs in under a millisecond for all sizes, whereas PennyLane slows dramatically, making \ours over \textbf{260,000$\times$ faster}.}
    \label{fig:runtime_speedup_t}
\end{figure}

Figure~\ref{fig:runtime_speedup_t} shows that PennyLane's runtime increases quickly as the circuit size grows, taking over \textbf{47 seconds} for circuits with 290 gates. In contrast, \ours completes the preprocessing in less than 200 microseconds in all cases. This shows that Clifford elimination is an essential and time-consuming step in current FTQC frameworks. PennyLane's backend is written in \textbf{C++}, so the large speed difference is due to the different algorithms used: PennyLane uses a DAG-based approach, while \ours uses a Tableau-based method.

By significantly reducing this preprocessing time, \ours makes Clifford elimination almost instantaneous, removing a major obstacle for efficient large-scale FTQC optimization.

\subsection{\ours Scalability}

\begin{figure}[h!]
    \centering
    \includegraphics[width=\linewidth]{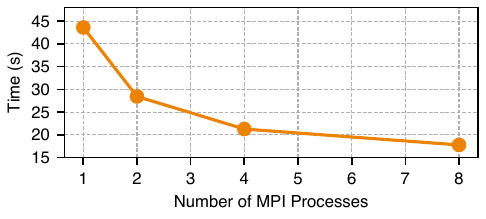}
    \caption{Scalability of \ours Clifford optimization using MPI on a 500-qubit QFT circuit with 4.8 million gates. Increasing from 1 to 8 processes yields a 2.45$\times$ speedup.}
    \label{fig:runtime_speedup_pennylane}
\end{figure}

To demonstrate the scalability of \ours, we evaluate its MPI-based parallel Clifford optimization (Section~\ref{sec:parallel_pbc}) on a 500-qubit QFT circuit with 4,776,699 gates. Figure~\ref{fig:runtime_speedup_pennylane} shows the scaling results from 1 to 8 processes. On a single process, \ours completes the optimization in 43.6 seconds. With 8 processes, the runtime decreases to 17.7 seconds, achieving a \textbf{2.45$\times$} speedup. This parallel gain compounds the already $10^5\times$ advantage of \ours over PennyLane, demonstrating strong scalability for large FTQC circuits.

\section{Related Works}

\subsection{Lattice Surgery Compiler}
Lattice surgery provides a measurement-based method for implementing logical gates in surface codes through patch merge and split operations, avoiding braiding while maintaining local connectivity \cite{horsman2012surface}. Modern lattice surgery compilers translate high-level Clifford+T circuits into surgery programs with explicit layouts and resource estimates \cite{watkins2024high}. The ZX-calculus offers a natural foundation, as lattice surgery operations correspond directly to ZX spider fusion and separation, enabling correctness-preserving rewrites that map to surgery primitives \cite{de2020zx}. Complementary design work explores cost trade-offs and surgery-friendly layouts for Clifford and non-Clifford operations, providing efficient strategies for handling T gates and magic-state distillation \cite{litinski2019game,litinski2018lattice}. These efforts establish lattice surgery not only as a practical compilation target but also as a bridge between abstract circuit identities and low-level fault-tolerant implementations. Importantly, lattice surgery compilation mainly aims to translate logical operations into low-level surgery operations, which is orthogonal to our circuit-level optimization that still requires lattice surgery compilation downstream.  

\subsection{General Quantum Circuit Optimization}

Quantum circuit optimization focuses on two main tracks: Noisy Intermediate-Scale Quantum(NISQ) and FTQC. NISQ optimization minimizes circuit depth for noise mitigation, while FTQC optimization targets logical operation efficiency.

\noindent\textbf{NISQ optimization.} For near-term devices, the primary limitation is the low fidelity of two-qubit gates. Optimizations, therefore, target reducing two-qubit gate count, depth, and hardware mapping costs. Techniques include peephole and template matching on Clifford circuits~\cite{bravyi2021clifford,iten2022exact}, combining rewriting and unitary synthesis for general circuit optimization~\cite{xu2025optimizing}, and compiler pipelines such as Qiskit that integrate layout, routing, and calibration-aware scheduling~\cite{qiskit2024,qiskitruntime}. 

\noindent\textbf{FTQC optimization.} In the fault-tolerant regime, the dominant cost arises from T gates, as each requires expensive magic-state distillation. Optimizations thus focus on minimizing T count and T depth. Foundational approaches include polynomial-time T-depth reduction via matroid partitioning~\cite{amy2014polynomial}, T-count minimization with Reed--Muller decoding~\cite{amy2019t}, and ZX-calculus--based reductions~\cite{kissinger2020reducing}. However, recent work~\cite{van2023optimising} shows that such T-optimization is generally hard and often incurs high runtime overhead. Other efforts optimize magic-state distillation protocols~\cite{fowler2018low,litinski2019magic} or explore magic-state cultivation techniques~\cite{gidney2024magic}, which suggest T gates may be realized at near-Clifford cost. 

These FTQC optimizations are orthogonal to \ours and can be applied before \ours to improve performance further. More importantly, the efficient runtime of \ours avoids the scaling issues that affect most FTQC optimization methods.

\subsection{System Optimization of FTQC}

Recent research has addressed critical system-level challenges in fault-tolerant quantum computing (FTQC) architectures. These advances include methods for synchronizing error-correction operations~\cite{maurya2025synchronization}, designing efficient qubit layouts for surface codes on neutral-atom platforms~\cite{viszlai2025interleaved}, and implementing speculative decoding techniques to meet stringent latency requirements~\cite{viszlai2025swiper}. 

Significant progress has also been made in optimizing lattice-surgery synthesis for logical operations~\cite{tan2024sat} and developing heterogeneous FTQC architectures that combine surface codes with quantum low-density parity-check codes for enhanced resource efficiency~\cite{stein2025hetec}. Further decoder optimizations~\cite{higgott2022pymatching,Das2022,das2022afs,das2022lilliput,Vittal2023,higgott2023sparseblossom,alavisamani2024,lee2025color} have also been explored.

\section{Conclusion}
Fault-tolerant quantum computing (FTQC) demands a large number of physical qubits, making resource-efficient designs essential when physical qubits are limited. However, these architectures introduce constraints that are often hidden at the logical circuit level, where two circuits with nearly identical metrics can have drastically different execution times. \ours addresses this by explicitly incorporating hardware constraints into circuit optimization, achieving a 2.57$\times$ (up to 12.2$\times$) improvement in FTQC runtime. 

Furthermore, FTQC optimization is inherently challenging, and without carefully designed algorithms, even effective techniques can be impractical. We tackle this by using a tableau-based approach instead of the DAGs employed in most other circuit optimization frameworks. Existing tableau designs are optimized for Clifford simulation and are not well-suited for circuit compilation. To address this, we develop specialized tableau variants tailored for optimization tasks. This brings over 1000$\times$ speedup compared to DAG-based methods, making \ours both highly efficient and scalable to large FTQC circuits. With tableaus at the heart of the stabilizer formalism, our optimized design lays the groundwork for future FTQC circuit improvements.

\begin{acks}
This material is based upon work supported by the U.S. Department of Energy, Office of Science, National Quantum Information Science Research Centers, Quantum Science Center. This research used resources of the Oak Ridge Leadership Computing Facility, which is a DOE Office of Science User Facility supported under Contract DE-AC05-00OR22725. This research used resources of the National Energy Research Scientific Computing Center (NERSC), a U.S. Department of Energy Office of Science User Facility located at Lawrence Berkeley National Laboratory, operated under Contract No. DE-AC02-05CH11231. The Pacific Northwest National Laboratory is operated by Battelle for the U.S. Department of Energy under Contract DE-AC05-76RL01830. In addition, this work was partially supported by the National Research Council (NRC) Canada through grants AQC 003 and 213, as well as by the Natural Sciences and Engineering Research Council of Canada (NSERC) under grant RGPIN-2019-05059.

\end{acks}

\balance
\bibliographystyle{ACM-Reference-Format}
\bibliography{ref}

\end{document}